\documentclass[preprintnumbers,twocolumn,prl,amsmath,floatfix]{revtex4}
\usepackage{pslatex}

\usepackage{graphicx,color}
\usepackage{dcolumn}

\begin{document}

\title{Intercalation and High Temperature Superconductivity of
Fullerides II}
\author{A.~Bill}
\email{abill@psi.ch}
\affiliation{Paul Scherrer Institute, Condensed Matter Theory, 5232
Villigen PSI, Switzerland}

\author{V.Z.~Kresin}
\affiliation{Lawrence Berkeley Laboratory, University of California at
        Berkeley, CA 94720, USA}

\begin{abstract}
Based on the mechanism described by the authors in Ref.~\cite{bill},
we predict that the compound C$_{60}$/CHI$_3$ will have $T_c \simeq
140$K, which would be the highest value obtained sofar for
intercalated materials.
\end{abstract}

\maketitle

This paper is based on our recent analysis \cite{bill} of high-temperature
superconductivity observed in intercalated fullerides \cite{schoen1}
(see also Ref.~\cite{service}). Following the approach presented in
Ref.~\cite{bill} we predict that the compound C$_{60}$/CHI$_3$ has a
value of $T_c$ which is presently the highest for intercalated
superconductors and second after the high-$T_c$ cuprates.

The authors of Ref.~\cite{schoen1} have discovered recently that
by intercalating CHCl$_3$ into the hole-doped C$_{60}$
fulleride (via field-effect doping, see Ref.~\cite{schoen2}) 
one obtains a critical temperature of $T_c \simeq 80$K.
Intercalating CHBr$_3$ molecule leads to a further increase to
$T_c \simeq 117$K(!).

It is believed that the superconducting state of fullerides is caused
by the coupling of the carriers with intramolecular vibrations of the
C$_{60}$ units (see e.g.~Ref.~\cite{gunnarsson}). With the
field-effect technique \cite{schoen2} it is possible to dope the
parent compound with either electrons or holes. In the latter case the
critical temperature of the fulleride ($T_c\simeq 52$K) is higher than for
usual n-doped materials, because the hole band has a larger density of
states \cite{schoen2}.

\begin{figure}[htp]
\includegraphics[width=5cm]{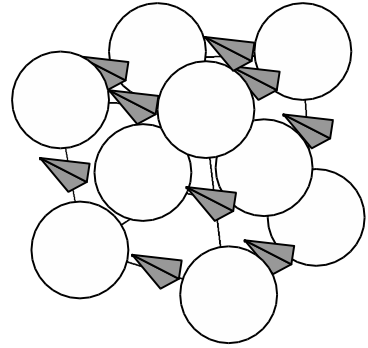}
\caption{\label{Fig1} Schematic representation of the C$_{60}$/CH$A_3$
  compound. The spheres represent the C$_{60}$ molecules arranged on an
  fcc lattice (only those from the visible faces are depicted for
  clarity). The tetrahedra represent the CH$A_3$ molecules
  (cf.~Fig.~\ref{Fig2}). The orientation of the molecules indicate
  that the actual lattice is hexagonal \cite{jansen}.}
\end{figure}

\begin{figure}[ht]
\begin{picture}(7,6)
\includegraphics[width=6cm]{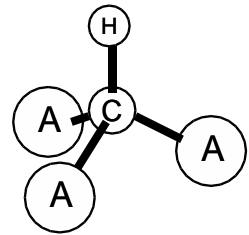}
\end{picture}
\vspace*{-2.5cm}
\caption{\label{Fig2} CH$A_3$ molecule where $A\equiv$Cl,Br,I. These
  molecules are intercalated into the hole-doped C$_{60}$ parent
  compound (represented as tetrahedra in Fig.~\ref{Fig1}.}
\end{figure}

The presence of intercalated molecules in the hole-doped parent
compound leads to a drastic increase in
$T_c$ ($52$K$\to 80$K$\to 117$K). As described in
Ref.~\cite{bill} we think that this increase is caused by the additional
interaction of the carriers with the vibrational manifold of the
intercalated molecules CH$A_3$ ($A\equiv$Cl,Br). The mechanism caused
by the addition of polyatomic molecules and consequently of new phonon
modes, was introduced by one of the authors in Ref.~\cite{kresin}. The
beauty of fullerides is that their molecular structure allows to
perform the intercalation (see Fig.~\ref{Fig1}).
\unitlength1cm
Our analysis \cite{bill} was based on the vibrational spectra of
CHCl$_3$ and CHBr$_3$ molecules, (Fig.~\ref{Fig2}; see
\cite{herzberg}). The increase in $T_c$ caused by the CHCl$_3 \to$
CHBr$_3$ substitution is due to the ``softening mechanism'' (see, e.g.,
Ref.~\cite{McMillan1}), that is to the decrease in characteristic
vibrational frequencies and corresponding increase in the coupling
constant $\lambda$ \cite{McMillan1}. Note that in the system under
consideration the characteristic mass entering the
expression of the coupling constant is $M\simeq M_C$, where $M_C$ is
the mass of the Carbon ion \cite{bill}.

We propose here to intercalate another molecule in the parent
fulleride (see Ref.~\cite{schoen2}), namely iodoform. That is, we
suggest to make the compound C$_{60}$/CHI$_3$. The CHI$_3$ molecule
has the same structure as CH$A_3$ ($A\equiv$Cl,Br;
cf.~Fig.~\ref{Fig2}).
\unitlength1cm
It has similar five vibrational modes (see
Ref.~\cite{iod}): $\Omega^{(1)I} = 110$cm$^{-1}$, $\Omega^{(2)I} =
154$cm$^{-1}$, $\Omega^{(3)I} = 425$cm$^{-1}$, $\Omega^{(4)I} =
578$cm$^{-1}$, $\Omega^{(1)I} = 1068$cm$^{-1}$ (the notation is
that of Ref.~\cite{bill}). One can compare these values with those
for CHCl$_3$ and CHBr$_3$ and directly see the additional "softening"
occuring in CHI$_3$. We have determined in Ref.~\cite{bill}, the
additional coupling constants (relative to the parent compound) for
the C$_{60}$/CHCl$_3$ and C$_{60}$/CHBr$_3$ compounds and found the
values $\lambda_2^{Cl}\simeq 0.2$ and $\lambda_2^{Br}\simeq
0.55$ \footnote{The value of $\mu^\star$ used here and in Ref.~\cite{bill}
is $\mu^\star\simeq 0.15$; note the misprint in Ref.~\cite{bill}}. Because
of the  softening effect, the coupling constant for the
C$_{60}$/CHI$_3$ compound is even larger and equals to
$\lambda_2^{I}\simeq 1.1$. As a result, we obtain $T_c\simeq
140$K. The nature of the approximations \cite{bill} is such that this
value is a lower bound for the $T_c$ of the C$_{60}$/CHI$_3$ compound.

The structure of the CHI$_3$ molecule is similar to that of CH$A_3$
($A\equiv$Cl,Br), and the intercalation thus seems realistic from the
point of view of chemistry. Since the $T_c$ obtained theoretically is
the highest for this class of materials it would be very interesting
to synthesize the C$_{60}$/CHI$_3$ compound and measure its
$T_c$.

\end{document}